\documentstyle[osa,manuscript]{revtex}

\input epsf.sty 
\input epsfig.sty

\begin{document}

\thispagestyle{empty}
\renewcommand{\thefootnote}{\fnsymbol{footnote}}


{\*\vspace{2.0cm}} 

\begin{center}
{\bf\Large Renormalization Group Reduction of Non Integrable 
Hamiltonian Systems} 

\vspace{1cm} 

Stephan I. Tzenov \\ 
{\it Plasma Physics Laboratory, Princeton University, Princeton, 
New Jersey 08543} 
\end{center} 

\medskip 

\vfill

\renewcommand{\baselinestretch}{1}
\normalsize

\begin{center}
{\bf\large   
Abstract }
\end{center}

\begin{quote}
Based on the Renormalization Group method, a reduction of 
non integrable multi-dimensional hamiltonian systems has 
been performed. The evolution equations for the slowly 
varying part of the angle-averaged phase space density, 
and for the amplitudes of the angular modes have been 
derived. It has been shown that these equations are 
precisely the Renormalization Group equations. As an 
application of the approach developed, the modulational 
diffusion in one-and-a-half degree of freedom dynamical 
system has been studied in detail. 
\end{quote}

\vfill

\begin{center} 
{\it Submitted to the} {\bf New Journal of Physics} 
\end{center}

\newpage


\renewcommand{\baselinestretch}{1}
\normalsize


%
\pagestyle{plain}

\renewcommand{\theequation}{\thesection.\arabic{equation}}

\setcounter{equation}{0}

\section{Introduction}

It is well-known that dynamical systems may exhibit irregular motion 
in certain regions of phase space \cite{Chirikov,Lichtenberg}. These 
regions differ in size, from being considerably small, to occupying 
large parts of phase space. This depends mostly on the strength of 
the perturbation, as well as on the intrinsic characteristics of the 
system. For comparatively small perturbations the regularity of the 
motion is expressed in the existence of adiabatic action invariants. 
In the course of nonlinear interaction the action invariants vary 
within a certain range, prescribed by the integrals of motion (if 
such exist). For chaotic systems some (or all) of the integrals of 
motion are destroyed, causing specific trajectories to become 
extremely complicated. These trajectories look random in their 
behavior, therefore it is natural to explore the statistical 
properties of chaotic dynamical systems.

Much experimental and theoretical evidence 
\cite{Chattopadhyay1,Chattopadhyay2} of nonlinear effects in the 
dynamics of particles in accelerators and storage rings is available 
at present. An individual particle propagating in an accelerator 
experiences growth of amplitude of betatron oscillations in a plane 
transverse to the particle orbit, whenever a perturbing force acts 
on it. This force may be of various origin, for instance high order 
multipole magnetic field errors, space charge forces, beam-beam 
interaction force, power supply ripple or other external and 
collective forces. Therefore, the Hamiltonian governing the motion 
of a beam particle is far from being integrable, and an irregular 
behavior of the beam is clearly observed, especially for a large 
number of turns. 

The idea to treat the evolution of chaotic dynamical systems in a 
statistical sense is not new; many rigorous results related to the 
statistical properties of such systems can be found in the book by 
Arnold and Avez \cite{Arnold}. Many of the details concerning the 
transport phenomena in the space of adiabatic action invariants only 
are also well understood \cite{Lichtenberg}. In this aspect the 
results presented here are in a sense re-derivation of previously 
obtained ones by means of a different method. What is new however, 
is the approach followed to obtain the diffusion properties in action 
variable space, as well as a new evolution equation for the 
angle-dependent part of the phase space density. Furthermore, instead 
of the widely used phenomenological method to derive the diffusion 
coefficient (tensor), the procedure pursued in the present paper is 
more consistent one, with a starting point the Liouville's equation 
for the phase space density.

We first employ the Projection Operator method of Zwanzig 
\cite{Zwanzig} to derive the equations for the two parts of the phase 
space density: the averaged over the angle variables part $F$, and 
the remainder $G$ [see Eq. (\ref{Projected}) in the next Section]. As 
expected, the two equations are coupled. Next we extract the relevant 
long-time scale behavior embedded in the equations for $F$ and $G$ 
by means of the Renormalization Group (RG) method \cite{Chen,Nozaki}. 
It is remarkable, and at the same time not surprising that the 
equations governing the long-time scale dynamics are the 
Renormalization Group equations (RGEs). These are obtained in 
Section 4 through renormalizing the perturbative solution of the 
equations for $F$ and $G$ [see Eqs. (\ref{Fdecomp}) and 
(\ref{Gdecomp}) of Section 2]. Finally, in Section 5 one-dimensional 
example of a chaotic system is considered to demonstrate the approach 
developed here. 

\renewcommand{\theequation}{\thesection.\arabic{equation}}

\setcounter{equation}{0}

\section{Projection Operator Method}

Single particle dynamics in cyclic accelerators and storage rings is 
most properly described by the adiabatic action invariants 
(Courant-Snyder invariants \cite{Courant}) and the canonically 
conjugate to them angle variables. However, to be more general we 
consider here a dynamical system with $N$ degrees of freedom, 
governed by the Hamiltonian written in action-angle variables 
${\left( {\bbox J}, {\bbox \alpha} \right)}$ as 
\begin{eqnarray} 
H{\left( {\bbox \alpha}, {\bbox J}; \theta \right)} = 
H_0{\left( {\bbox J} \right)} + \epsilon 
V{\left( {\bbox \alpha}, {\bbox J}; \theta \right)}, 
\label{Hamiltonian} 
\end{eqnarray} 
\noindent 
where $\theta$ is the independent azimuthal variable (widely used in 
accelerator physics), playing the role of time and ${\bbox J}$ and 
${\bbox \alpha}$ are $N$-dimensional vectors
\begin{eqnarray} 
{\bbox J} = {\left( J_1, J_2, \dots, J_N \right)}, 
\qquad \qquad \qquad 
{\bbox \alpha} = {\left( \alpha_1, \alpha_2, \dots, 
\alpha_N \right)}. 
\label{Actangle} 
\end{eqnarray} 
\noindent 
Moreover $H_0{\left( {\bbox J} \right)}$ is the integrable part of 
the Hamiltonian, $\epsilon$ is a formal small parameter, while 
$V{\left( {\bbox \alpha}, {\bbox J}; \theta \right)}$ is the 
perturbation periodic in the angle variables 
\begin{eqnarray} 
V{\left( {\bbox \alpha}, {\bbox J}; \theta \right)} = 
{\sum \limits_{\bbox m}}^{\prime} 
V_{\bbox m}{\left( {\bbox J}; \theta \right)} 
\exp {\left( i {\bbox m} \cdot {\bbox \alpha} \right)}, 
\label{Perturbat} 
\end{eqnarray} 
\noindent 
where $\sum'$ denotes exclusion of the harmonic 
${\bbox m} = (0, 0, \dots, 0)$ from the above sum. The Hamilton's 
equations of motion are 
\begin{eqnarray} 
{\frac {{\rm d} \alpha_k} {{\rm d} \theta}} = 
\omega_{0k}{\left( {\bbox J} \right)} + 
\epsilon {\frac {\partial V} {\partial J_k}}, 
\qquad \qquad \qquad \qquad 
{\frac {{\rm d} J_k} {{\rm d} \theta}} = - 
\epsilon {\frac {\partial V} {\partial \alpha_k}}, 
\label{Hamilton} 
\end{eqnarray} 
\noindent 
where 
\begin{eqnarray} 
\omega_{0k}{\left( {\bbox J} \right)} = 
{\frac {\partial H_0} {\partial J_k}}. 
\label{Tune} 
\end{eqnarray} 
\noindent 
In what follows (in particular in Section 4) we assume that the 
nonlinearity coefficients 
\begin{eqnarray} 
\gamma_{kl}{\left( {\bbox J} \right)} = 
{\frac {\partial^2 H_0} {\partial J_k \partial J_l}} 
\label{Nonlinearity} 
\end{eqnarray} 
\noindent 
are small and can be neglected. The Liouville's equation governing 
the evolution of the phase space density 
$P{\left( {\bbox \alpha}, {\bbox J}; \theta \right)}$ can be written 
as 
\begin{eqnarray} 
{\frac {\partial} {\partial \theta}} 
P{\left( {\bbox \alpha}, {\bbox J}; \theta \right)} = 
{\left[ {\widehat{\cal L}}_0 + \epsilon 
{\widehat{\cal L}}_v (\theta) \right]} 
P{\left( {\bbox \alpha}, {\bbox J}; \theta \right)}. 
\label{Liouville} 
\end{eqnarray} 
\noindent 
Here the operators ${\widehat{\cal L}}_0$ and ${\widehat{\cal L}}_v$ 
are given by the expressions 
\begin{eqnarray} 
{\widehat{\cal L}}_0 = - 
\omega_{0k}{\left( {\bbox J} \right)} 
{\frac {\partial} {\partial \alpha_k}}, 
\qquad \qquad \qquad \qquad 
{\widehat{\cal L}}_v = 
{\frac {\partial V} {\partial \alpha_k}} 
{\frac {\partial} {\partial J_k}} - 
{\frac {\partial V} {\partial J_k}} 
{\frac {\partial} {\partial \alpha_k}}, 
\label{Operators} 
\end{eqnarray} 
\noindent 
where summation over repeated indices is implied. Next we define the 
projection operator onto the subspace of action variables by the 
following integral: 
\begin{eqnarray} 
{\widehat{\cal P}} f{\left( {\bbox J}; \theta \right)} = 
{\frac {1} {(2 \pi)^N}} \int \limits_0^{2 \pi} 
{\rm d} \alpha_1 \dots \int \limits_0^{2 \pi} 
{\rm d} \alpha_N 
f{\left( {\bbox \alpha}, {\bbox J}; \theta \right)}, 
\label{Projection} 
\end{eqnarray} 
\noindent 
where $f{\left( {\bbox \alpha}, {\bbox J}; \theta \right)}$ is a 
generic function of its arguments. Let us introduce also the functions 
\begin{eqnarray} 
F = {\widehat{\cal P}} P, \qquad \qquad \qquad 
G = {\left( 1 - {\widehat{\cal P}} \right)} P = 
{\widehat{\cal C}} P, \qquad \qquad \qquad 
{\left( P = F + G \right)}. 
\label{Projected} 
\end{eqnarray} 
\noindent 
From Eq. (\ref{Liouville}) with the obvious relations 
\begin{eqnarray} 
{\widehat{\cal P}} {\widehat{\cal L}}_0 = 
{\widehat{\cal L}}_0 {\widehat{\cal P}} \equiv 0, 
\qquad \qquad \qquad \qquad \qquad 
{\widehat{\cal P}} {\widehat{\cal L}}_v 
{\widehat{\cal P}} \equiv 0 
\label{Relations} 
\end{eqnarray} 
\noindent 
in hand it is straightforward to obtain the equations 
\begin{eqnarray} 
{\frac {\partial F} {\partial \theta}} = 
\epsilon {\widehat{\cal P}} {\widehat{\cal L}}_v G = 
\epsilon {\frac {\partial} {\partial J_k}} 
{\widehat{\cal P}} {\left( 
{\frac {\partial V} {\partial \alpha_k}} G 
\right)}, 
\label{Fequation} 
\end{eqnarray} 
\begin{eqnarray} 
{\frac {\partial G} {\partial \theta}} = 
{\widehat{\cal L}}_0 G + \epsilon 
{\widehat{\cal C}} {\widehat{\cal L}}_v G + 
\epsilon {\widehat{\cal L}}_v F. 
\label{Gequation} 
\end{eqnarray} 

Our goal in the subsequent exposition is to analyze Eqs. 
(\ref{Fequation}) and (\ref{Gequation}) using the RG method. It will 
prove efficient to eliminate the dependence on the angle variables 
in $G$ and $V$ by noting that the eigenfunctions of the operator 
${\widehat{\cal L}}_0$ form a complete set, so that every function 
periodic in the angle variables can be expanded in this basis. Using 
Dirac's ``bra-ket'' notation we write 
\begin{eqnarray} 
{\left| {\bbox n} \right \rangle} = 
{\frac {1} {(2 \pi)^{N/2}}} \exp 
{\left( i {\bbox n} \cdot {\bbox \alpha} \right)}, 
\qquad \qquad \qquad 
{\left \langle {\bbox n} \right|} = 
{\frac {1} {(2 \pi)^{N/2}}} \exp 
{\left( - i {\bbox n} \cdot {\bbox \alpha} \right)}. 
\label{Eigenfunc} 
\end{eqnarray} 
\noindent 
The projection operator ${\widehat{\cal P}}$ can be represented in 
the form \cite{Sudbery} 
\begin{eqnarray} 
{\widehat{\cal P}} = {\widehat{\cal P}}_{\bbox 0} = 
{\left| {\bbox 0} \right \rangle} 
{\left \langle {\bbox 0} \right|}. 
\label{Project} 
\end{eqnarray} 
\noindent 
One can also define a set of projection operators 
${\widehat{\cal P}}_{\bbox n}$ according to the expression 
\cite{Sudbery} 
\begin{eqnarray} 
{\widehat{\cal P}}_{\bbox n} = 
{\left| {\bbox n} \right \rangle} 
{\left \langle {\bbox n} \right|}. 
\label{Projectn} 
\end{eqnarray} 
\noindent 
It is easy to check the completeness relation 
\begin{eqnarray} 
\sum \limits_{\bbox n} 
{\widehat{\cal P}}_{\bbox n} = 1, 
\label{Complete} 
\end{eqnarray} 
\noindent 
from which and from Eq. (\ref{Project}) it follows that 
\begin{eqnarray} 
{\widehat{\cal C}} = 
{\sum \limits_{{\bbox n} \neq {\bbox 0}}}^{\prime} 
{\left| {\bbox n} \right \rangle} 
{\left \langle {\bbox n} \right|}. 
\label{Coperator} 
\end{eqnarray} 
Decomposing the quantities $F$, $G$ and $V$ in the basis 
(\ref{Eigenfunc}) as 
\begin{eqnarray} 
F = F{\left( {\bbox J}; \theta \right)} 
{\left| {\bbox 0} \right \rangle}, 
\label{Ffunc} 
\end{eqnarray} 
\begin{eqnarray} 
G{\left( {\bbox \alpha}, {\bbox J}; \theta \right)} = 
{\sum \limits_{{\bbox m} \neq {\bbox 0}}}^{\prime} 
G_{\bbox m}{\left( {\bbox J}; \theta \right)} 
{\left| {\bbox m} \right \rangle}, 
\label{Gfunc} 
\end{eqnarray} 
\begin{eqnarray} 
V{\left( {\bbox \alpha}, {\bbox J}; \theta \right)} = 
{\sum \limits_{{\bbox n} \neq {\bbox 0}}}^{\prime} 
V_{\bbox n}{\left( {\bbox J}; \theta \right)} 
{\left| {\bbox n} \right \rangle}, 
\label{Vfunc} 
\end{eqnarray} 
\noindent 
from Eqs. (\ref{Fequation}) and (\ref{Gequation}) we obtain 
\begin{eqnarray} 
{\frac {\partial F} {\partial \theta}} = 
i \epsilon {\frac {\partial} {\partial J_k}} 
{\left( {\sum \limits_{\bbox n}}^{\prime} 
n_k V_{\bbox n} G_{-{\bbox n}} \right)}, 
\label{Fdecomp} 
\end{eqnarray} 
\begin{eqnarray} 
{\frac {\partial G_{\bbox n}} {\partial \theta}} = 
- i n_k \omega_{0k} G_{\bbox n} + 
i \epsilon {\sum \limits_{\bbox m}}^{\prime} 
{\left[ n_k V_{{\bbox n}-{\bbox m}} 
{\frac {\partial G_{\bbox m}} {\partial J_k}} - 
m_k {\frac {\partial} {\partial J_k}} 
{\left( V_{{\bbox n}-{\bbox m}} G_{\bbox m} \right)} 
\right]} + i \epsilon n_k V_{\bbox n} 
{\frac {\partial F} {\partial J_k}}. 
\label{Gdecomp} 
\end{eqnarray} 

The above equations comprise the the starting point in the analysis 
outlined in Section 4. We are primarily interested in the long-time 
evolution of the original system governed by certain amplitude 
equations. These will turn out to be precisely the RG equations. 

\renewcommand{\theequation}{\thesection.\arabic{equation}}

\setcounter{equation}{0}

\section{Renormalization Group Reduction of Hamilton's Equations} 

Let us consider the solution of Hamilton's equations of motion 
(\ref{Hamilton}) for small perturbation parameter $\epsilon$. It is 
natural to introduce the perturbation expansion 
\begin{eqnarray} 
\alpha_k = \alpha_k^{(0)} + \epsilon 
\alpha_k^{(1)} + \epsilon^2 \alpha_k^{(2)} 
+ \cdots, 
\qquad \qquad \qquad 
J_k = J_k^{(0)} + \epsilon J_k^{(1)} + 
\epsilon^2 J_k^{(2)} + \cdots. 
\label{Acanexpand} 
\end{eqnarray} 
\noindent 
The lowest order perturbation equations have the trivial solution: 
\begin{eqnarray} 
\alpha_k^{(0)} = \omega_{0k} \theta 
+ \varphi_k, 
\qquad \qquad \qquad 
J_k^{(0)} = A_k, 
\label{Acansol0} 
\end{eqnarray} 
\noindent 
where $A_k$ and $\varphi_k$ are constant amplitude and phase, 
respectively. We write the first order perturbation equations as 
\begin{eqnarray} 
{\frac {{\rm d} \alpha_k^{(1)}} {{\rm d} \theta}} = 
\gamma_{kl}{\left( {\bbox A} \right)} J_l^{(1)} + 
{\frac {\partial V} {\partial A_k}}, 
\qquad \qquad \qquad \qquad 
{\frac {{\rm d} J_k^{(1)}} {{\rm d} \theta}} = - 
{\frac {\partial V} {\partial \alpha_k^{(0)}}}, 
\label{Hamil1} 
\end{eqnarray} 
\noindent 
Assuming that the modes 
$V_{\bbox n}{\left( {\bbox J}; \theta \right)}$ are periodic in 
$\theta$, we can expand them in a Fourier series 
\begin{eqnarray} 
V_{\bbox n}{\left( {\bbox J}; \theta \right)} = 
\sum \limits_{\mu = - \infty}^{\infty} 
V_{\bbox n}{\left( {\bbox J}; \mu \right)} 
\exp {\left( i \mu \nu_{\bbox n} \theta \right)}. 
\label{Modes} 
\end{eqnarray} 
\noindent 
If the original system (\ref{Hamiltonian}) is far from primary 
resonances of the form: 
\begin{eqnarray} 
n_k^{(R)} \omega_{0k} + \mu \nu_R = 0 
\label{Resonances} 
\end{eqnarray} 
\noindent 
we can solve the first order perturbation equations (\ref{Hamil1}), 
yielding the result: 
\begin{eqnarray} 
\alpha_k^{(1)} &=& i {\sum \limits_{\bbox m}}^{\prime} 
{\sum \limits_{\mu}} \gamma_{kl} 
{\left( {\bbox A} \right)} m_l V_{\bbox m} (\mu) 
{\frac {\exp{\left[ i {\left( m_s \omega_{0s} + \mu 
\nu_{\bbox m}  \right)} \theta \right]}} 
{{\left( m_s \omega_{0s} + \mu 
\nu_{\bbox m} \right)}^2}} 
{\exp{\left( i m_s \varphi_s \right)}} 
\nonumber \\ 
&-& i {\sum \limits_{\bbox m}}^{\prime} 
{\sum \limits_{\mu}} 
{\frac {\partial V_{\bbox m} (\mu)} {\partial A_k}} 
{\frac {\exp{\left[ i {\left( m_s \omega_{0s} + \mu 
\nu_{\bbox m}  \right)} \theta \right]}} 
{m_s \omega_{0s} + \mu \nu_{\bbox m}}} 
{\exp{\left( i m_s \varphi_s \right)}}, 
\label{Ansol1} 
\end{eqnarray} 
\begin{eqnarray} 
J_k^{(1)} = - {\sum \limits_{\bbox m}}^{\prime} 
{\sum \limits_{\mu}} m_k  V_{\bbox m} (\mu) 
{\frac {\exp{\left[ i {\left( m_s \omega_{0s} + \mu 
\nu_{\bbox m}  \right)} \theta \right]}} 
{m_s \omega_{0s} + \mu \nu_{\bbox m}}} 
{\exp{\left( i m_s \varphi_s \right)}}, 
\label{Acsol1} 
\end{eqnarray} 
\noindent 
The second order perturbation equations have the form: 
\begin{eqnarray} 
{\frac {{\rm d} \alpha_k^{(2)}} {{\rm d} \theta}} = 
\gamma_{kl}{\left( {\bbox A} \right)} J_l^{(2)} + 
{\frac {1} {2}} 
{\frac {\partial \gamma_{kl}} {\partial A_s}} 
J_l^{(1)} J_s^{(1)} + 
{\frac {\partial^2 V} {\partial A_k \partial A_l}} 
J_l^{(1)} + {\frac {\partial^2 V} 
{\partial A_k \partial \alpha_l^{(0)}}} 
\alpha_l^{(1)}, 
\label{Anhamil2} 
\end{eqnarray} 
\begin{eqnarray} 
{\frac {{\rm d} J_k^{(2)}} {{\rm d} \theta}} = - 
{\frac {\partial^2 V} 
{\partial \alpha_k^{(0)} \partial A_l}} 
J_l^{(1)} - {\frac {\partial^2 V} 
{\partial \alpha_k^{(0)} \partial \alpha_l^{(0)}}} 
\alpha_l^{(1)}. 
\label{Achamil2} 
\end{eqnarray} 
\noindent 
The solution to Eq. (\ref{Achamil2}) reads as 
\begin{eqnarray} 
J_k^{(2)} &=& 2 \pi {\cal R} (\theta) 
{\sum \limits_{{\bbox m}>{\bbox 0}}}^{\prime} 
{\sum \limits_{\mu}} m_k m_l 
{\frac {\partial {\left| V_{\bbox m} 
{\left( \mu \right)} \right|}^{\bbox 2}} 
{\partial A_l}} 
\Re {\left( \gamma; \; m_s \omega_{0s} + \mu 
\nu_{\bbox m} \right)} 
\nonumber \\ 
&+& 2 \pi {\cal R} (\theta) 
{\sum \limits_{{\bbox m}>{\bbox 0}}}^{\prime} 
{\sum \limits_{\mu}} m_k m_l m_s 
\gamma_{ls} {\left( {\bbox A} \right)} 
{\left| V_{\bbox m} 
{\left( \mu \right)} \right|}^{\bbox 2} 
{\left. {\frac {\partial} {\partial a}} 
\Re {\left( \gamma; \; a \right)} 
\right|}_{a = m_r \omega_{0r} + \mu 
\nu_{\bbox m}} 
\nonumber \\ 
&+& {\rm oscillating \; terms}, 
\label{Acsol2} 
\end{eqnarray} 
\noindent 
where 
\begin{eqnarray} 
{\frac {{\rm d} {\cal R}} 
{{\rm d} \theta}} = 1, 
\label{Protocoeff} 
\end{eqnarray} 
\begin{eqnarray} 
\pi \Re (x; \; y) = 
{\frac {x} {x^2 + y^2}}, 
\qquad \qquad \qquad \qquad 
\lim \limits_{x \rightarrow 0} 
\Re (x; \; y) = \delta (y), 
\label{Refunc} 
\end{eqnarray} 
\noindent 
and $\gamma$ is a small real quantity added {\it ad hoc} in the 
denominators of the expressions (\ref{Ansol1}) and (\ref{Acsol1}). 
The limit $\gamma \rightarrow 0$ will be taken in the final result. 

As expected, in the second order perturbation solution (\ref{Acsol2}) 
the first and the second terms are secular, because 
${\cal R}(\theta) = \theta$. To remove these secularities we follow 
the general prescription of the RG method \cite{Chen,Nozaki}. First, 
we select the slowly varying part of the perturbation solution 
governing the long-time evolution of the system. Up to second order 
in the perturbation parameter $\epsilon$ it consists of the constant 
zero order term $A_k$ and the second order secular terms. Next, we 
introduce the intermediate time $\tau$, and in order to absorb the 
difference $\tau = \theta - (\theta - \tau)$ we make the new 
renormalized amplitude $A_k (\tau)$ dependent on $\tau$. Since the 
long-time solution thus constructed should not depend on $\tau$ its 
derivative with respect to $\tau$ must be equal to zero. This also 
holds for $\tau = \theta$, so that finally 
\begin{eqnarray} 
{\frac {{\rm d} A_k (\theta)} {{\rm d} \theta}} 
&=& 2 \pi \epsilon^2 
{\sum \limits_{{\bbox m}>{\bbox 0}}}^{\prime} 
{\sum \limits_{\mu}} m_k m_l 
{\frac {\partial {\left| V_{\bbox m} 
{\left( {\bbox A}; \mu \right)} \right|}^{\bbox 2}} 
{\partial A_l}} \Re {\left( \gamma; \; m_s \omega_{0s} 
+ \mu \nu_{\bbox m} \right)} 
\nonumber \\ 
&+& 2 \pi \epsilon^2 
{\sum \limits_{{\bbox m}>{\bbox 0}}}^{\prime} 
{\sum \limits_{\mu}} m_k m_l m_s 
\gamma_{ls} {\left( {\bbox A} \right)} 
{\left| V_{\bbox m} 
{\left( {\bbox A}; \mu \right)} \right|}^{\bbox 2} 
{\left. {\frac {\partial} {\partial a}} 
\Re {\left( \gamma; \; a \right)} 
\right|}_{a = m_r \omega_{0r} + \mu 
\nu_{\bbox m}}. 
\label{Acrg} 
\end{eqnarray} 

Equation (\ref{Acrg}) is known as the Renormalization Group 
equation (RGE). It describes the slow long-time evolution of the 
action variables. 

\renewcommand{\theequation}{\thesection.\arabic{equation}}

\setcounter{equation}{0}

\section{Renormalization Group Reduction of Liouville's Equation} 

We consider the solution of Eqs. (\ref{Fdecomp}) and (\ref{Gdecomp}) 
for small $\epsilon$ by means of the RG method. For that purpose we 
perform again the naive perturbation expansion 
\begin{eqnarray} 
F = F^{(0)} + \epsilon  F^{(1)} + 
\epsilon^2  F^{(2)} + \cdots, 
\qquad \qquad \qquad 
G_{\bbox n} = G_{\bbox n}^{(0)} + 
\epsilon G_{\bbox n}^{(1)} + 
\epsilon^2 G_{\bbox n}^{(2)} + \cdots, 
\label{Fgexpand} 
\end{eqnarray} 
\noindent 
and substitute it in Eqs. (\ref{Fdecomp}) and (\ref{Gdecomp}). The 
lowest order perturbation equations have the obvious solution 
\begin{eqnarray} 
F^{(0)} = F_0 {\left( {\bbox J} \right)}, 
\qquad \qquad \qquad \qquad 
G_{\bbox n}^{(0)} = W_{\bbox n} 
{\left( {\bbox J} \right)} 
\exp {\left( -i n_k \omega_{0k} \theta \right)}. 
\label{Zsolution} 
\end{eqnarray} 
\noindent 
The first order perturbation equations read as: 
\begin{eqnarray} 
{\frac {\partial F^{(1)}} {\partial \theta}} = 
i {\frac {\partial} {\partial J_k}} 
{\left[ {\sum \limits_{\bbox n}}^{\prime} 
n_k V_{\bbox n} W_{-{\bbox n}} 
\exp {\left( i n_l \omega_{0l} \theta \right)}
\right]}, 
\label{Ffequat} 
\end{eqnarray} 
\begin{eqnarray} 
{\frac {\partial G_{\bbox n}^{(1)}} {\partial \theta}} = 
- i n_k \omega_{0k} G_{\bbox n}^{(1)} + 
i n_k V_{\bbox n} {\frac {\partial F_0} {\partial J_k}} 
\nonumber 
\end{eqnarray} 
\begin{eqnarray} 
+ i {\sum \limits_{\bbox m}}^{\prime} 
{\left[ n_k V_{{\bbox n}-{\bbox m}} 
{\frac {\partial W_{\bbox m}} {\partial J_k}} - 
m_k {\frac {\partial} {\partial J_k}} 
{\left( V_{{\bbox n}-{\bbox m}} W_{\bbox m} \right)} 
\right]} \exp {\left( -i  m_l \omega_{0l} \theta \right)}. 
\label{Fgequat} 
\end{eqnarray} 
\noindent 
We again assume that the modes 
$V_{\bbox n}{\left( {\bbox J}; \theta \right)}$ are periodic in 
$\theta$, so that they can be expanded in a Fourier series 
(\ref{Modes}). If the original system (\ref{Hamiltonian}) 
exhibits primary resonances of the form (\ref{Resonances}) 
in the case when $\omega_{0k}$ does not depend on the action 
variables, we can solve the first order perturbation equations 
(\ref{Ffequat}) and (\ref{Fgequat}). The result is as follows: 
\begin{eqnarray} 
F^{(1)} &=& i {\cal R}(\theta) 
{\sum \limits_{{\bbox n}^{(R)}}}^{\prime \prime} 
n_k^{(R)} {\frac {\partial} {\partial J_k}} 
{\left[ V_{{\bbox n}^{(R)}} {\left( 
- {\frac {n_l^{(R)} \omega_{0l}} {\nu_R}} 
\right)} W_{-{\bbox n}^{(R)}} \right]} \nonumber \\ 
&+& {\sum \limits_{\bbox n}}^{\prime} 
{\sum \limits_{\mu}}^{\prime} n_k 
{\frac {\partial} {\partial J_k}} {\left[ 
V_{\bbox n} (\mu) W_{-{\bbox n}} \right]} 
{\frac {\exp {\left[ i {\left( 
n_l \omega_{0l} + \mu \nu_{\bbox n} 
\right)} \theta \right]}} 
{n_l \omega_{0l} + \mu \nu_{\bbox n}}}, 
\label{Ffsolut} 
\end{eqnarray} 
\begin{eqnarray} 
G_{\bbox n}^{(1)} = {\cal G}_{\bbox n} 
\exp {\left( -i n_k \omega_{0k} 
\theta \right)}, 
\label{Fgsolut} 
\end{eqnarray} 
\noindent 
where 
\begin{eqnarray} 
{\cal G}_{\bbox n} = i {\cal R}(\theta) 
\delta_{{\bbox n} {\bbox n}^{(R)}} n_k 
{\frac {\partial F_0} {\partial J_k}} 
V_{\bbox n} {\left( 
- {\frac {n_l^{(R)} \omega_{0l}} {\nu_R}} 
\right)} 
\nonumber 
\end{eqnarray} 
\begin{eqnarray} 
+ i {\cal R}(\theta) 
{\sum \limits_{{\bbox n}-{\bbox n}^{(R)}}}^{\prime \prime} 
{\left\{ n_k V_{{\bbox n}^{(R)}} {\left( 
- {\frac {n_l^{(R)} \omega_{0l}} {\nu_R}} 
\right)} {\frac {\partial W_{{\bbox n}-{\bbox n}^{(R)}}} 
{\partial J_k}} \right.} 
\nonumber 
\end{eqnarray} 
\begin{eqnarray} 
{\left. - {\left( n_k - n_k^{(R)} \right)} 
{\frac {\partial} {\partial J_k}} 
{\left[ V_{{\bbox n}^{(R)}} {\left( 
- {\frac {n_l^{(R)} \omega_{0l}} {\nu_R}} 
\right)} W_{{\bbox n}-{\bbox n}^{(R)}} 
\right]} \right\}} 
\nonumber 
\end{eqnarray} 
\begin{eqnarray} 
+ n_k {\frac {\partial F_0} {\partial J_k}} 
{\sum \limits_{\mu}}^{\prime} 
V_{\bbox n} (\mu) 
{\frac {\exp {\left[ i {\left( 
n_l \omega_{0l} + \mu \nu_{\bbox n} 
\right)} \theta \right]}} 
{n_l \omega_{0l} + \mu \nu_{\bbox n}}} 
\nonumber 
\end{eqnarray} 
\begin{eqnarray} 
+ {\sum \limits_{\bbox m}}^{\prime} 
{\sum \limits_{\mu}}^{\prime} 
{\left\{ n_k V_{{\bbox n}-{\bbox m}} (\mu) 
{\frac {\partial W_{\bbox m}} {\partial J_k}} 
- m_k {\frac {\partial} {\partial J_k}} 
{\left[ V_{{\bbox n}-{\bbox m}} (\mu) 
 W_{\bbox m} \right]} \right\}} 
{\frac {\exp {\left\{ i {\left[ 
{\left( n_l - m_l \right)} 
\omega_{0l} + \mu \nu_{{\bbox n}-{\bbox m}} 
\right]} \theta \right\}}} 
{{\left( n_l - m_l \right)} \omega_{0l} + 
\mu \nu_{{\bbox n}-{\bbox m}}}} 
\label{Calg} 
\end{eqnarray} 
\noindent 
In the above expressions $\sum''$ denotes summation over all 
primary resonances (\ref{Resonances}). To obtain the desired 
RG equations we proceed in the same way as in the previous Section. 
The first order RG equations are 
\begin{eqnarray} 
{\frac {\partial F_0} {\partial \theta}} = 
i \epsilon 
{\sum \limits_{{\bbox n}^{(R)}}}^{\prime \prime} 
n_k^{(R)} {\frac {\partial} {\partial J_k}} 
{\left[ V_{{\bbox n}^{(R)}} {\left( {\bbox J}; 
\mu_R \right)} W_{-{\bbox n}^{(R)}} 
{\left( {\bbox J} \right)} \right]}, 
\label{Rgfequat} 
\end{eqnarray} 
\begin{eqnarray} 
{\frac {\partial W_{\bbox n}} {\partial \theta}} = 
i \epsilon \delta_{{\bbox n} {\bbox n}^{(R)}} n_k 
{\frac {\partial F_0} {\partial J_k}} 
V_{\bbox n} {\left( {\bbox J}; \mu_R \right)} 
\nonumber 
\end{eqnarray} 
\begin{eqnarray} 
+ i \epsilon 
{\sum \limits_{{\bbox n}-{\bbox n}^{(R)}}}^{\prime \prime} 
{\left\{ n_k V_{{\bbox n}^{(R)}} {\left( {\bbox J}; \mu_R 
\right)} {\frac {\partial W_{{\bbox n}-{\bbox n}^{(R)}}} 
{\partial J_k}} - {\left( n_k - n_k^{(R)} \right)} 
{\frac {\partial} {\partial J_k}} 
{\left[ V_{{\bbox n}^{(R)}} {\left( {\bbox J}; \mu_R 
\right)} W_{{\bbox n}-{\bbox n}^{(R)}} 
{\left( {\bbox J} \right)} \right]} \right\}}, 
\label{Rgwequat} 
\end{eqnarray} 
\noindent 
where 
\begin{eqnarray} 
\mu_R = - 
{\frac {n_k^{(R)} \omega_{0k}} {\nu_R}}. 
\label{Mur} 
\end{eqnarray} 
\noindent 
Equations (\ref{Rgfequat}) and (\ref{Rgwequat}) describe the 
resonant mode coupling when strong primary resonances are present 
in the original system. 

Let us now assume that the original system is far from resonances. 
Solving the second order perturbation equation for $F^{(2)}$ and 
$G_{\bbox n}^{(2)}$
\begin{eqnarray} 
{\frac {\partial F^{(2)}} {\partial \theta}} = 
i {\frac {\partial} {\partial J_k}} 
{\left( {\sum \limits_{\bbox n}}^{\prime} 
n_k V_{\bbox n} G_{-{\bbox n}}^{(1)} \right)}, 
\label{Sfequat} 
\end{eqnarray} 
\begin{eqnarray} 
{\frac {\partial G_{\bbox n}^{(2)}} {\partial \theta}} + 
i n_k \omega_{0k} G_{\bbox n}^{(2)} = 
i n_k V_{\bbox n} {\frac {\partial F^{(1)}} {\partial J_k}} 
+ i {\sum \limits_{\bbox m}}^{\prime} 
{\left[ n_k V_{{\bbox n}-{\bbox m}} 
{\frac {\partial G_{\bbox m}^{(1)}} {\partial J_k}} - 
m_k {\frac {\partial} {\partial J_k}} 
{\left( V_{{\bbox n}-{\bbox m}} G_{\bbox m}^{(1)} \right)} 
\right]}, 
\label{Sgequat} 
\end{eqnarray} 
\noindent 
we obtain 
\begin{eqnarray} 
F^{(2)} = 2 \pi {\cal R} (\theta) 
{\frac {\partial} {\partial J_k}} {\left[ 
{\sum \limits_{{\bbox n}>{\bbox 0}}}^{\prime} 
{\sum \limits_{\lambda}} 
n_k n_l {\left| V_{\bbox n} 
{\left( \lambda \right)} \right|}^{\bbox 2} 
{\frac {\partial F_0} {\partial J_l}} \Re 
{\left( \gamma; \; n_s \omega_{0s} + \lambda 
\nu_{\bbox n} \right)} \right]} + {\rm oscillating \; terms}, 
\label{Sfsolut} 
\end{eqnarray} 
\begin{eqnarray} 
G_{\bbox n}^{(2)} = {\cal F}_{\bbox n} 
\exp {\left( - i n_s \omega_{0s} \theta \right)}, 
\label{Sgsolut} 
\end{eqnarray} 
\noindent 
where 
\begin{eqnarray} 
{\cal F}_{\bbox n} = i {\cal R} (\theta) 
n_k n_l {\sum \limits_{\mu}} 
{\frac {V_{\bbox n} (\mu)} 
{n_s \omega_{0s} + \mu \nu_{\bbox n}}}
{\frac {\partial^2} {\partial J_k \partial J_l}} 
{\left[ V_{\bbox n}^{\ast} (\mu) W_{\bbox n} \right]} 
\nonumber 
\end{eqnarray} 
\begin{eqnarray} 
+ i {\cal R} (\theta) 
{\sum \limits_{\bbox m}}^{\prime} 
{\sum \limits_{\mu}} {\frac {1} 
{{\left( n_s - m_s \right)} \omega_{0s} + 
\mu \nu_{{\bbox n}-{\bbox m}}}} 
{\left\{ - n_k m_l V_{{\bbox n}-{\bbox m}} (\mu) 
{\frac {\partial} {\partial J_k}} {\left( 
V_{{\bbox n}-{\bbox m}}^{\ast} (\mu) 
{\frac {\partial W_{\bbox n}} {\partial J_l}} 
\right)} \right.} 
\nonumber 
\end{eqnarray} 
\begin{eqnarray} 
+ n_k n_l V_{{\bbox n}-{\bbox m}} (\mu) 
{\frac {\partial^2} {\partial J_k \partial J_l}} 
{\left( V_{{\bbox n}-{\bbox m}}^{\ast} (\mu) 
W_{\bbox n} \right)} + m_k m_l 
{\frac {\partial} {\partial J_k}} {\left( 
{\left| V_{{\bbox n}-{\bbox m}} (\mu) \right|}^{\bf 2} 
{\frac {\partial W_{\bbox n}} {\partial J_l}} \right)} 
\nonumber 
\end{eqnarray} 
\begin{eqnarray} 
{\left. - m_k n_l {\frac {\partial} {\partial J_k}} 
{\left[ V_{{\bbox n}-{\bbox m}} (\mu) 
{\frac {\partial} {\partial J_l}} {\left( 
V_{{\bbox n}-{\bbox m}}^{\ast} (\mu) 
W_{\bbox n} \right)} \right]} \right\}} 
+ {\rm oscillating \; terms}, 
\label{Scfsolut} 
\end{eqnarray} 
\noindent 
and the functions ${\cal R} (\theta)$ and $\Re (x; \; y)$ are given 
by Eqs. (\ref{Protocoeff}) and (\ref{Refunc}), respectively. It is 
now straightforward to write the second order Renormalization 
Group equations. They are: 
\begin{eqnarray} 
{\frac {\partial F_0} {\partial \theta}} = 
2 \pi \epsilon^2 
{\frac {\partial} {\partial J_k}} {\left[ 
{\sum \limits_{{\bbox n}>{\bbox 0}}}^{\prime} 
{\sum \limits_{\lambda}} 
n_k n_l {\left| V_{\bbox n} 
{\left( \lambda \right)} \right|}^{\bbox 2} 
{\frac {\partial F_0} {\partial J_l}} \Re 
{\left( \gamma; \; n_s \omega_{0s} + \lambda 
\nu_{\bbox n} \right)} \right]}, 
\label{Frgequat} 
\end{eqnarray} 
\begin{eqnarray} 
{\frac {\partial W_{\bbox n}} {\partial \theta}} = 
i \epsilon^2 
n_k n_l {\sum \limits_{\mu}} 
{\frac {V_{\bbox n} (\mu)} 
{n_s \omega_{0s} + \mu \nu_{\bbox n}}}
{\frac {\partial^2} {\partial J_k \partial J_l}} 
{\left[ V_{\bbox n}^{\ast} (\mu) W_{\bbox n} \right]} 
\nonumber 
\end{eqnarray} 
\begin{eqnarray} 
+ i \epsilon^2 
{\sum \limits_{\bbox m}}^{\prime} 
{\sum \limits_{\mu}} {\frac {1} 
{{\left( n_s - m_s \right)} \omega_{0s} + 
\mu \nu_{{\bbox n}-{\bbox m}}}} 
{\left\{ - n_k m_l V_{{\bbox n}-{\bbox m}} (\mu) 
{\frac {\partial} {\partial J_k}} {\left( 
V_{{\bbox n}-{\bbox m}}^{\ast} (\mu) 
{\frac {\partial W_{\bbox n}} {\partial J_l}} 
\right)} \right.} 
\nonumber 
\end{eqnarray} 
\begin{eqnarray} 
+ n_k n_l V_{{\bbox n}-{\bbox m}} (\mu) 
{\frac {\partial^2} {\partial J_k \partial J_l}} 
{\left( V_{{\bbox n}-{\bbox m}}^{\ast} (\mu) 
W_{\bbox n} \right)} + m_k m_l 
{\frac {\partial} {\partial J_k}} {\left( 
{\left| V_{{\bbox n}-{\bbox m}} (\mu) \right|}^{\bf 2} 
{\frac {\partial W_{\bbox n}} {\partial J_l}} \right)} 
\nonumber 
\end{eqnarray} 
\begin{eqnarray} 
{\left. - m_k n_l {\frac {\partial} {\partial J_k}} 
{\left[ V_{{\bbox n}-{\bbox m}} (\mu) 
{\frac {\partial} {\partial J_l}} {\left( 
V_{{\bbox n}-{\bbox m}}^{\ast} (\mu) 
W_{\bbox n} \right)} \right]} \right\}}. 
\label{Wrgequat} 
\end{eqnarray} 

The RG equation (\ref{Frgequat}) is a Fokker-Planck equation describing 
the diffusion of the adiabatic action invariant. It has been derived 
previously by many authors (see e.g. the book by Lichtenberg and 
Lieberman \cite{Lichtenberg} and the references therein). It is important 
to note that our derivation does not require the initial assumption 
concerning the fast stochastization of the angle variable. The fact that 
the latter is indeed a stochastic variable is clearly visible from the 
second RG equation (\ref{Wrgequat}), governing the slow amplitude 
evolution of the angle-dependent part of the phase space density. 
Nevertheless it looks complicated, its most important feature is that 
equations for the amplitudes of different modes are decoupled. In the 
case of isolated nonlinear resonance Eq. (\ref{Wrgequat}) acquires a 
very simple form as will be shown in the next Section. 

\renewcommand{\theequation}{\thesection.\arabic{equation}}

\setcounter{equation}{0}

\section{Modulational Diffusion}

As an example to demonstrate the theory developed in previous sections, 
we consider the simplest example of one-and-a-half degree of freedom 
dynamical system exhibiting chaotic motion 
\begin{eqnarray} 
H_0 {\left( J \right)} = 
\lambda J + H_s {\left( J \right)}, 
\qquad \qquad \qquad 
V {\left( \alpha, J; \theta \right)} = 
V {\left( J \right)} \cos 
{\left( \alpha + \xi \sin \nu \theta 
\right)}. 
\label{Onedham} 
\end{eqnarray} 
\noindent 
The Hamiltonian (\ref{Onedham}), written in resonant canonical 
variables describes an isolated nonlinear resonance of one-dimensional 
betatron motion of particles in an accelerator with modulated resonant 
phase (or modulated linear betatron tune). The modulation may come from 
various sources: ripple in the power supply of quadrupole magnets, 
synchro-betatron coupling or ground motion. The resonance detuning 
$\lambda$ defines the distance from the resonance, $\Xi$ is the 
amplitude of modulation of the linear betatron tune and $\nu$ is the 
modulation frequency, where $\xi = \Xi / \nu$. Without loss of 
generality we consider $\xi$ positive. Since
\begin{eqnarray} 
\omega_0 = \lambda + 
{\frac {{\rm d} H_s} {{\rm d} J}}, 
\qquad \qquad \qquad 
V_1 {\left( J; \mu \right)} = 
{\frac {1} {2}} V {\left( J \right)} 
{\cal J}_{\mu} {\left( \xi \right)}, 
\label{Onedmode} 
\end{eqnarray} 
\noindent 
where ${\cal J}_n {\left( z \right)}$ is the Bessel function of 
order $n$, the RG equation (\ref{Acrg}) for the amplitude $A$ can be 
rewritten as 
\begin{eqnarray} 
{\frac {{\rm d} A} {{\rm d} \theta}} = 
{\frac {\pi \epsilon^2} {2 \nu}} {\left\{ 
{\frac {\partial} {\partial A}} {\left[ 
V^2 {\left( A \right)} 
{\cal J}_{\left[ {\frac {\omega_0} {\nu}} \right]}^2 
{\left( \xi \right)} 
\right]} - \gamma {\left( A \right)} 
V^2 {\left( A \right)} {\left. 
{\frac {\partial} {\partial a}} {\left[ 
{\cal J}_a^2 {\left( \xi \right)} 
\right]} \right|}_{a = - {\frac {\omega_0} {\nu}}}
\right\}}. 
\label{Onedrgex} 
\end{eqnarray} 
\noindent 
Here the square brackets $[z]$ encountered in the index of the Bessel 
function imply integer part of $z$. Moreover, in deriving the 
expression for $V_1 {\left( J; \mu \right)}$ in Eq. (\ref{Onedmode}) 
use has been made of the identity 
\begin{eqnarray} 
{\exp {\left( i q \sin z \right)}} = 
= \sum \limits_{n=- \infty}^{\infty} 
{\cal J}_n {\left( {\left| q \right|} \right)} 
\exp {\left[ i n z {\rm sgn} 
{\left( q \right)} \right]}, 
\label{Bessel} 
\end{eqnarray} 
\noindent 
and finally, the limit $\gamma \rightarrow 0$ in Eq. (\ref{Acrg}) has 
been taken. For small value of $\xi$ utilizing the approximate 
expression for the derivative of Bessel functions with respect to the 
order we obtain 
\begin{eqnarray} 
{\frac {{\rm d} A} {{\rm d} \theta}} = 
{\frac {\pi \epsilon^2} {2 \nu}} {\left\{ 
{\frac {\partial} {\partial A}} {\left[ 
V^2 {\left( A \right)} 
{\cal J}_{\left[ {\frac {\omega_0} {\nu}} \right]}^2 
{\left( \xi \right)} 
\right]} - 2 
{\cal J}_{\left[ {\frac {\omega_0} {\nu}} \right]}^2 
{\left( \xi \right)} \ln {\left( 
{\frac {\xi} {2}} \right)} 
\gamma {\left( A \right)} 
V^2 {\left( A \right)} \right\}}. 
\label{Onedrgap} 
\end{eqnarray} 

\begin{figure} 
\centerline{\epsfxsize=15cm \epsfbox{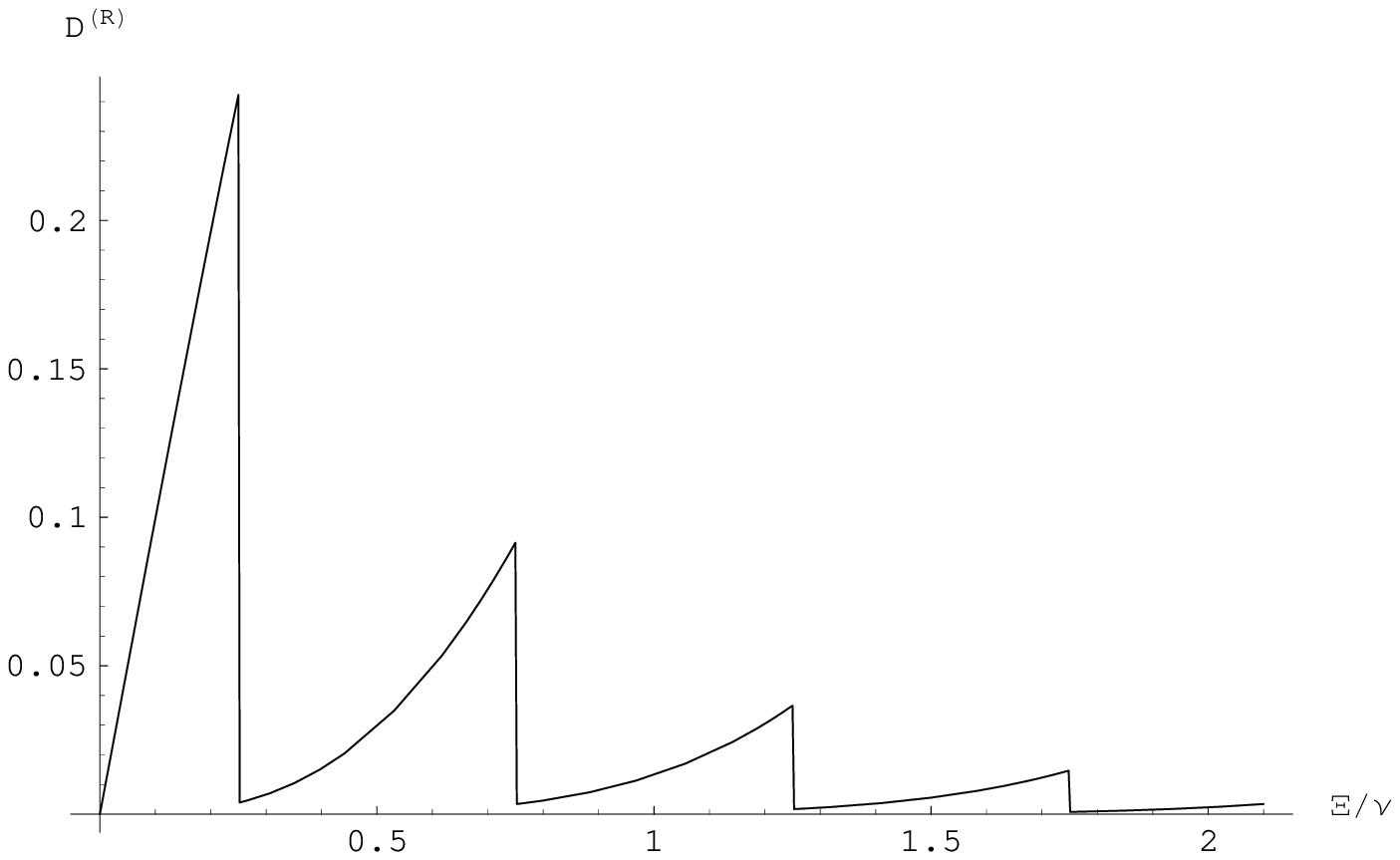}}
\caption{Reduced diffusion coefficient ${\cal D}^{(R)}$ as a function 
of the ratio $\xi = \Xi / \nu$ for $\lambda = 2 \Xi$.} 
\label{Fig1} 
\end{figure}

\begin{figure} 
\centerline{\epsfxsize=15cm \epsfbox{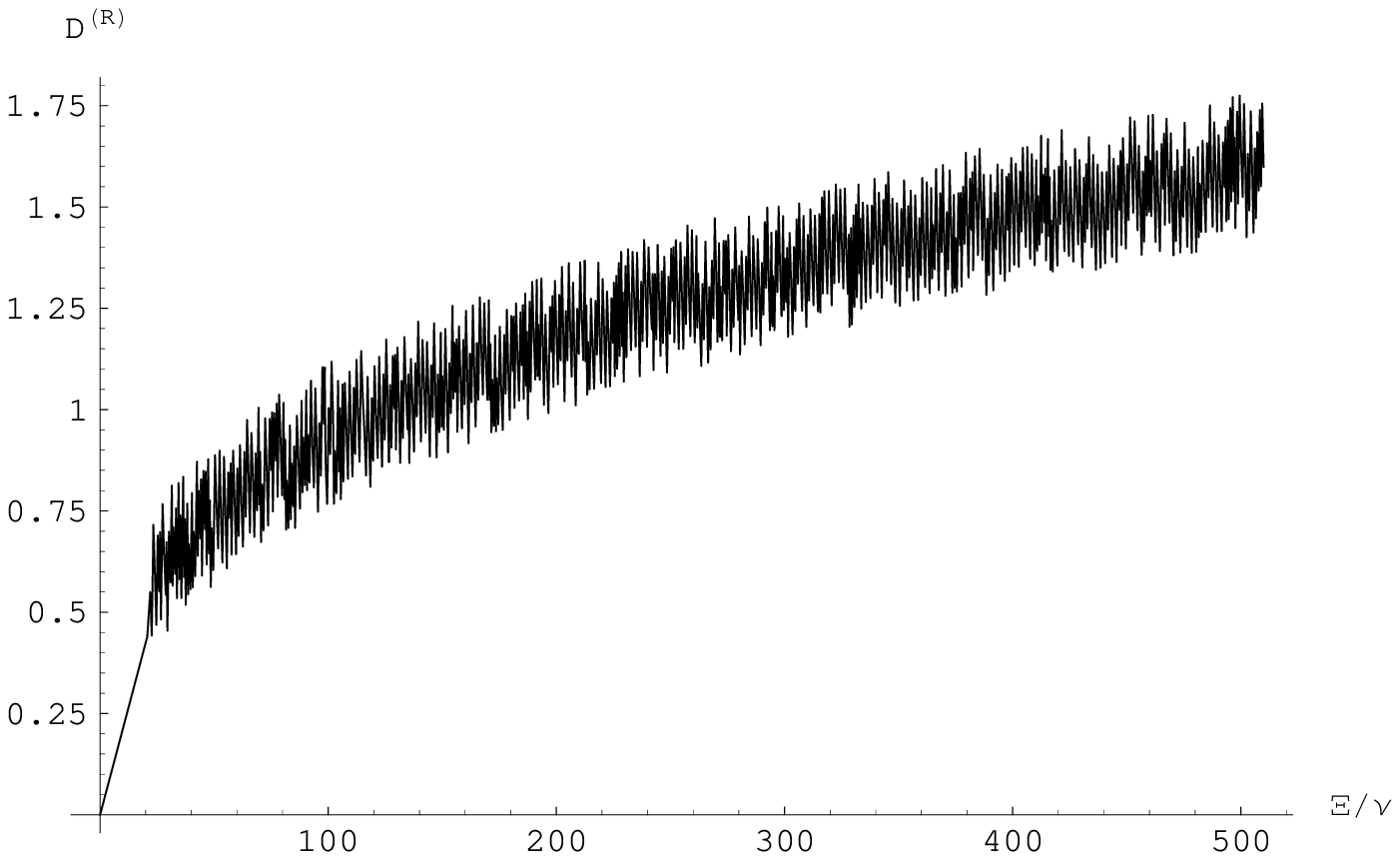}}
\caption{Reduced diffusion coefficient ${\cal D}^{(R)}$ as a function 
of the ratio $\xi = \Xi / \nu$ for $\lambda = \Xi$.} 
\label{Fig2} 
\end{figure}

\begin{figure} 
\centerline{\epsfxsize=15cm \epsfbox{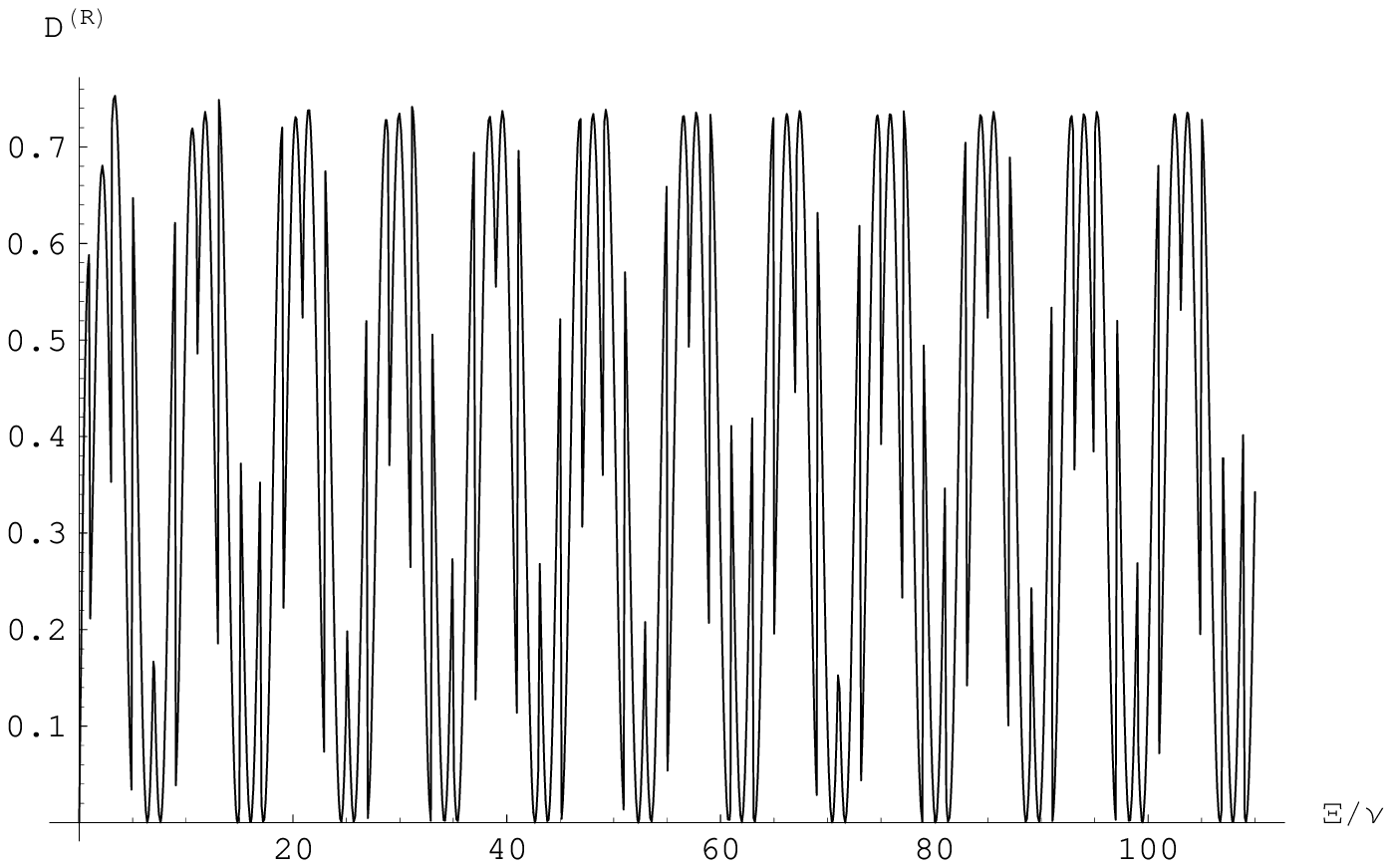}}
\caption{Reduced diffusion coefficient ${\cal D}^{(R)}$ as a function 
of the ratio $\xi = \Xi / \nu$ for $\lambda = \Xi / 2$..} 
\label{Fig3} 
\end{figure}

Let us now turn to the RG equations (\ref{Frgequat}) and 
(\ref{Wrgequat}). They can be rewritten in the form: 
\begin{eqnarray} 
{\frac {\partial F_0} {\partial \theta}} = 
{\frac {\pi \epsilon^2} {2 \nu}} 
{\frac {\partial} {\partial J}} {\left[ 
V^2 {\left( J \right)} 
{\cal J}_{\left[ {\frac {\omega_0} {\nu}} \right]}^2 
{\left( \xi \right)} 
{\frac {\partial F_0} {\partial J}} \right]}, 
\label{Rgequatf0} 
\end{eqnarray} 
\begin{eqnarray} 
{\frac {1} {W_n}} 
{\frac {\partial W_n} {\partial \theta}} = 
{\frac {i \pi \epsilon^2 n} 
{2 \nu \sin {\left( \pi \omega_0 / \nu \right)}}} 
{\cal J}_{- {\frac {\omega_0} {\nu}}}^2 
{\left( \xi \right)} 
{\frac {{\rm d}} {{\rm d} J}} {\left( 
V {\frac {{\rm d} V} {{\rm d} J}} \right)} + 
{\frac {\pi \epsilon^2 n} {2 \nu}} 
{\cal J}_{\left[ {\frac {\omega_0} {\nu}} \right]}^2 
{\left( \xi \right)} 
{\frac {{\rm d}} {{\rm d} J}} {\left( 
V {\frac {{\rm d} V} {{\rm d} J}} \right)}. 
\label{Rgequatwn} 
\end{eqnarray} 
\noindent 
Equation (\ref{Rgequatwn}) suggests that the amplitudes $W_n$ of 
the angular modes $G_n$ exhibit exponential growth with an increment 
\begin{eqnarray} 
\Gamma = {\frac {\pi \epsilon^2} {2 \nu}} 
{\cal J}_{\left[ {\frac {\omega_0} {\nu}} \right]}^2 
{\left( \xi \right)} 
{\frac {{\rm d}} {{\rm d} J}} {\left( 
V {\frac {{\rm d} V} {{\rm d} J}} \right)}. 
\label{Increment} 
\end{eqnarray} 
\noindent 
Equation (\ref{Rgequatf0}) is a Fokker-Planck equation for the 
angle-independent part of the phase space density with a diffusion 
coefficient 
\begin{eqnarray} 
{\cal D} {\left( J \right)} = 
{\frac {\pi \epsilon^2} {2 \nu}} 
V^2 {\left( J \right)} 
{\cal J}_{\left[ {\frac {\omega_0} {\nu}} \right]}^2 
{\left( \xi \right)}. 
\label{Difcoeff} 
\end{eqnarray} 
\noindent 
In Figures \ref{Fig1}--\ref{Fig3} the reduced diffusion coefficient 
\begin{eqnarray} 
{\cal D}^{(R)} 
{\left( J, {\frac {\Xi} {\nu}} \right)} = 
{\frac {2 \Xi} {\pi \epsilon^2 V^2 {\left( J \right)}}} 
{\cal D} {\left( J \right)} 
\label{Redifcoeff} 
\end{eqnarray} 
\noindent 
has been plotted as a function of the ratio between the amplitude 
and the frequency of the modulation. Three typical regimes 
corresponding to different values of $\lambda / \Xi$ used as a 
control parameter have been chosen. In the first one depicted in 
Figure \ref{Fig1} we have taken the resonance detuning twice as 
large as the amplitude of the modulation 
${\left( \lambda = 2 \Xi \right)}$. In this case there is no 
crossing of the main resonance described by the Hamiltonian 
(\ref{Onedham}) and the diffusion coefficient decreases very 
rapidly after reaching its maximum value at $\xi = 0.25$. The 
cases of periodic resonance crossings for $\lambda = \Xi$ and 
$\lambda = \Xi / 2$ are shown in Figure \ref{Fig2} and Figure 
\ref{Fig3}, respectively. 

\renewcommand{\theequation}{\thesection.\arabic{equation}}

\setcounter{equation}{0}

\section{Concluding Remarks}

In the present paper we apply the Renormalization Group method to 
the reduction of non integrable multi-dimensional Hamiltonian 
systems. The notion of reduction is used here in the sense of 
slow, long-time behavior, that survives appropriate averaging and/or 
factorizing of rapidly oscillating contributions to the dynamics 
of the system. 

As a result of the investigation performed we have derived evolution 
equations for the slowly varying part of the angle-averaged phase 
space density, and for the amplitudes of the angular modes. It has 
been shown that these equations are the Renormalization Group 
equations. 

The case of a one-dimensional isolated nonlinear resonance with a 
resonant phase (or linear unperturbed tune) subjected to periodic 
modulation has been studied in detail. The coefficient of 
modulational diffusion, as well as the exponential growth increment 
of the amplitudes of angular modes have been obtained in explicit 
form. 

\subsection*{Acknowledgments}

It is a pleasure to thank E. Startsev for many helpful discussions 
concerning the subject of the present paper. I am indebted to N. 
Goldenfeld for careful reading of the manuscript, and for making 
valuable suggestions. This work was supported by the U.S. Department 
of Energy. 




\end{document}